\documentclass[fleqn,usenatbib]{mnras}
\usepackage[english]{babel}

\usepackage[T1]{fontenc}
\usepackage{ae,aecompl}

\usepackage{fancyhdr}
\usepackage{amsfonts}
\usepackage{amsmath}
\usepackage{amssymb}
\usepackage{multicol}
\usepackage{layout}
\usepackage{graphicx}
\usepackage{multirow}
\usepackage{color}
\usepackage{multirow}

\usepackage{times}
\usepackage{natbib}
\def\be{\begin{equation}}
\def\ee{\end{equation}}

\usepackage[outdir=./fig/]{epstopdf}

\newif\ifAMStwofonts
\AMStwofontstrue

\graphicspath{{./fig/}}

\title[Structure formation in PADE parameterization]{Structure formation in dark energy cosmologies described by PADE parameterization}

\author[Mehdi Rezaei]{Mehdi Rezaei $^{1,2}$\thanks{rezaei@irimo.ir}
\\
{$^1$ Research Institute for Astronomy and Astrophysics of Maragha (RIAAM)
, Maragha, Iran, P.O.Box:55134-441}\\
$^2$ Iran meteorological organization, Hamedan Research Center for Applied
Meteorology, Hamedan, Iran}

\date{Accepted ?, Received ?; in original form \today}

\pagerange{\pageref{firstpage}--\pageref{lastpage}} \pubyear{2015}

\begin{document}

\label{firstpage}

\maketitle

\begin{abstract}
We study the imprints on the formation of cosmic structures of a particular class of dark energy parameterizations dubbed PADE parameterization. Here we investigate how dark energy can affect the growth of large scale structures of  the universe in the framework of spherical collapse model. The dynamics of
the spherical collapse of a dark matter halo depends on the properties of the dark energy model. We show that the properties of spherical collapse scenario are directly affected by the evolution of dark energy. We obtain the main parameters of spherical collapse for two different DE parameterizations in two different  approaches:
first the homogeneous DE approach, in which dark energy does not exhibit fluctuations on cluster scales and the other, clustered DE scenario in which, dark energy components inside the overdense region collapses similar to dark matter. Using the Sheth-Tormen mass function, we investigate the abundance of virialized halos in the framework of PADE parameterizations. Specifically, the present analysis shows that the number count of dark matter halos depends on the evolution of DE parameterizations and clustering properties of dark energy. Also we show that perturbations in phantom DE components enhance the growth of matter perturbations. This result were obtained in the literature for dark energy parameterizations that are phantom at all redshifts. But we obtained same results for dark energy parameterizations which are in phantom regime and enter in quintessence region at relatively low redshifts. We also show that in parameterizations under study, low mass halos were formed before massive halos.
\end{abstract}
\maketitle


\section{Introduction}
In the last two decades, increasingly large body of cosmological observations including those of type Ia 
supernova  (SnIa) \citep{Riess1998,Perlmutter1999,Kowalski2008},
cosmic microwave background (CMB)
\citep{Komatsu2009,Jarosik:2010iu,Komatsu2011,Ade:2015yua}, baryonic acoustic oscillation (BAO)
\citep{Tegmark:2003ud,Cole:2005sx,Eisenstein:2005su,Percival2010,Blake:2011rj,Reid:2012sw}, high redshift galaxy clusters \citep{Wang1998,Allen:2004cd}, high redshift galaxies \citep{Alcaniz:2003qy}, and weak gravitational lensing \citep{Benjamin:2007ys,Amendola:2007rr,Fu:2007qq} indicate that our universe is experiencing a period of cosmic acceleration. There is an important question about the cosmological dynamics; what is the cause and nature of this accelerated expansion. To answer this question, cosmologists follow two main approaches. Some accept general relativity (GR) and try to explain accelerated expansion by introducing a new cosmic fluid with sufficiently negative pressure dubbed dark energy (DE). Based on the latest cosmological observations, this unknown fluid occupies about $3/4$ of the total energy budget of the universe \citep{Ade:2015yua}. On the other hand, some believe that this acceleration reflects on the physics of gravity at cosmological scales. They are trying to modify general relativity to explain this acceleration. In this way they propose modified gravity models that some of them have been investigated in the literature, $f(R)$ gravity \citep{Buchdahl:1983zz},Randall-Sundrum
model \citep{Randall:1999vf}, DGP model  \citep{Dvali:2000hr}, modified DGP model \citep{Koyama:2006ef} and so forth. 

In this work to justify the current acceleration of the universe, among these two approaches, we follow the first one, DE. The cosmological constant $\Lambda$ with $w_{\rm \Lambda}=-1$ is the first and simplest candidate for DE. Although the standard $\Lambda$ cosmology is consistent with observations, but it suffers from two puzzles, the so-called fine-tuning and cosmic coincidence problems \citep{Weinberg1989,Sahni:1999gb,Carroll2001,Padmanabhan2003,Copeland:2006wr}. The $\Lambda$ problems encourage cosmologists to seek new DE models with time evolving energy density in order to solve the above cosmological problems or at least alleviate them. Some of these attempts led to new dynamical DE models with time evolving EoS parameter proposed widely in literature in recent years.  Ghost DE \citep{Veneziano1979,Witten1979,Kawarabayashi1980,Rosenzweig1980}, quintessence \citep{Caldwell:1997ii,Erickson:2001bq}, holographic DE models \citep{Horava2000,Thomas2002}, k-essence\citep{Armendariz2001}, chaplygin gas\citep{Kamenshchik:2001cp}, generalized chaplygin gas\citep{Bento:2002ps}, dilaton \citep{Gasperini2002,Arkani2004,Piazza2004}, phantom\citep{Caldwell2002}, quintom\citep{Elizalde:2004mq}  and etc are some of these models. Recently, many of these models were compared with different observational data sets. Some of these comparisons show that DE models are consistent with latest observational data, but nevertheless $\Lambda$ cosmology is more consistent model with the observations yet \citep{Mehrabi:2015kta,Malekjani:2016edh,Malekjani:2018qcz,Rezaei:2019roe}.
In studying the nature of DE and its dynamic a precise measurement of EoS parameter and its variation can led to some important results\citep{Copeland:2006wr,Frieman:2008sn,Weinberg:2012es,Amendola:2012ys}.
As a simple way to investigate the nature of DE we can develope a formalism whereby we can directly apply some parameterizations of its EoS. 
In literature, one can find different forms of parameterizations for the  EoS of DE \citep{Maor:2000jy,Chevallier:2000qy,Linder:2002et,Riess:2004nr,Seljak:2004xh,Bassett:2007aw}. There is no mathematical principle or fundamental physics behind most of these parameterizations. In this work we investigate two types of PADE parameterization ( see section \ref{sect:pade}), which from the mathematical point of view seems to be more stable in comparison with other parameterizations. Our parameterizations do not diverge and thus can be used at both small and high redshifts. 

DE not only can be the cause of the accelerated expansion of the universe, but also can affect the structure formation scenario in universe. It is believed that the large scale structures (LSS) in universe are constructed from gravitational collapse of primordial small density perturbations\citep{Gunn1972,Press1974,White1978,Peebles1993,Peacock1999,Peebles2003}. Initial seeds of these density perturbations are produced during inflation \citep{Guth1981,Linde1990}.
In order to study the evolution of these fluctuations we have a simple scenario, the spherical collapse model (SCM) which first introduced by \citep{Gunn1972}.
In this scenario, any of small density perturbations assume to be a spherical overdense region. While background of universe is expanding, because of self-gravity, the spherical overdense region expands slower in comparison with background. Therefore the density of spherical overdense region compare to background becomes more and more. At  turnaround redshift, $z_{\rm ta}$,  the overdense sphere decouples from the Hubble fluid and starts to collapse. In a certain radius (virial radius) at redshift $z_{\rm vir}$The collapsing sphere region reaches to a steady state and here after known as a virialized halo. SCM in DE cosmologies has been widely studied in several works\citep{Ryden1987,Subramanian2000,Ascasibar2004,Williams2004,Pace:2013pea}. It has been extended for different cosmological models\citep{Mota2004,Maor2005,Abramo2007,Schaefer:2007nf,Abramo2009a,Li2009,Pace2010,Pace2012}. In this paper we study the SCM in the presence of DE cosmologies which their EoS parameters are types of PADE parameterizations. For these cosmologies we predict the abundance of virialized halos. The paper is organized as following. In section \ref{sect:pade}, we first introduce the main ingredients of PADE parameterizations , then we investigate the background evolution of the universe. In section \ref{growth}, the basic equations which introduce the evolution of density perturbations in both linear and nonlinear regimes are presented. In section \ref{sec:mass} we obtain the predicted mass function and cluster number count in our parameterizations in both homogeneous and clustered DE scenarios.
Finally we  summarize our results and conclude in section\ref{conclude} .

\section{BACKGROUND EVOLUTION IN PADE PARAMETERIZATIONS}\label{sect:pade}

For a function $f(x)$, the PADE approximation of order $(m,n)$ is the ratio of two polynomials as below \citep{pade1892,baker96,Adachi:2011vu}
\begin{eqnarray}\label{padeO}
f(x)=\frac{a_0+a_1x+a_2x^2+...+a_nx^n}{b_0+b_1x+b_2x^2+...+b_nx^m}\;,
\end{eqnarray}
where $m$ and $n$ are positive integers and $a_{\rm i}$ and $b{\rm _i}$ are constants. 
Setting $b_{\rm i}=0$ for $i\ge 1$, this approximation reduces to well known Taylor expansion. 
In this paper we focus on two especial form of PADE parameterizations as follows \citep[see also][]{Wei:2013jya}.

\subsection{parameterization (1)}
Using Eq. (\ref{padeO}), we expand the EoS parameter $w_{\rm de}$ up to order $(1,1)$ with respect to $(1-a)$ as below \citep{Wei:2013jya}:
\begin{equation}\label{pade1}
w_{\rm de}(a)=\frac{w_0 + w_{1}(1-a)}{1+w_{2}(1-a)}\;.
\end{equation}
In terms of redshift $z$, one can write Eq. (\ref{pade1}) as

\begin{equation}\label{pade1b}
w_{\rm de}(z)=\frac{w_0+(w_0+w_1)z}{1+(1+w_{2})z}\;.
\end{equation}

Setting $w_2=0$ Eq. (\ref{pade1}) reduces to famous CPL parameterization. It is easy to see that in parameterization (1) unlike CPL parameterization, just by setting $w_2\neq 0$ we can avoid the divergence of the EoS parameter at $z=-1$. Using Eq. (\ref{pade1}) we find: 

\begin{eqnarray}
w_{\rm de}=\left\{
\begin{array}{ll}
\frac{w_0+w_1}{1+w_2}\,,~~ & {\rm at~early~time}~~
(a\to 0 ~{\rm or}~z\to\infty)\,,\\[4mm]
w_0\,, & {\rm at~present}~~ (a=1~{\rm or}~z=0)\,,\\[4mm]
\frac{w_1}{w_2}\,, & {\rm at~far~future}~~ 
(a\to\infty ~{\rm or}~z\to -1)\,,
\end{array} \right.
\end{eqnarray}
Setting $w_2\not=0$ and $w_2\not=-1$, parameterization (1) will be a well-behaved function at redshift interval $-1\leq z\leq\infty$.\\

In isotropic, homogeneous and spatially flat Friedmann-Robertson-Walker (FRW) cosmologies, the first Friedmann equation reads 
\begin{eqnarray}\label{frid1}
H^2=\frac{8\pi G}{3}(\rho_{\rm r}+\rho_{\rm m}+\rho_{\rm de})\;,
\end{eqnarray}
where $H\equiv {\dot a}/a$ is the Hubble parameter, $\rho_{\rm r}$ is the energy density of radiation and $\rho_{\rm m}$ and $\rho_{\rm de}$ are the relevant energy densities for dark matter and DE components respectively. In the absence of interactions between these components we have
 \begin{eqnarray}\label{continuity}
 && \dot{\rho_{\rm r}}+4H\rho_{\rm r}=0\;,\label{radiation}\\
&&\dot{\rho_{\rm m}}+3H\rho_{\rm m}=0\;,\label{matter}\\
&&\dot{\rho_{\rm de}}+3H\rho_{\rm de}(1+w_{\rm de})=0\;\label{de},
 \end{eqnarray}
where the over-dot displays derivative  with respect to cosmic time $t$. Inserting Eq . (\ref{pade1}) into Eq. (\ref{de}), one can obtain the DE density of parameterization (1) as
\citep[see also][]{Wei:2013jya}

\begin{eqnarray}
\rho_{\rm de}=\rho^{(0)}_{\rm de} a^{-3(\frac{1+w_0 + w_1+ w_2}{1+w_2})} [1+w_2 (1-a)]^{-3(\frac{w_1- w_0 w_2}{w_2 (1+w_2)})}\;,\label{rho-pade1}
\end{eqnarray}

Also, combining Eq.(\ref{rho-pade1}) and Eq.(\ref{frid1}) we derive the dimensionless Hubble 
parameter $(E=H/H_0)$ for  parameterization (1) as 

\begin{eqnarray}
&&E^{2}=\Omega_ {\rm r0} a^{-4}+\Omega_{\rm m0} a^{-3} + (1-[\Omega_{\rm r0}+\Omega_{\rm m0}]) \times \nonumber \\
&&a^{-3(\frac{1+w_0+ w_1 + w_2}{1+w_2})} \times (1+w_2 - a w_2)^{-3(\frac{w_1 -w_0 w_2}{w_2(1+w_2)})} \;,\label{Epade1}
\end{eqnarray}
where $\Omega_{\rm m0}, \Omega_{\rm r0}$ and $\Omega_{\rm de0}$ are density parameter, radiation parameter and DE parameter respectively and in spatially flat universe we have $\Omega_{\rm m0}+\Omega_{\rm r0}+\Omega_{\rm de0}=1$.

\subsection{parameterization (2)}
Clearly, parameterization (1) has three free parameters $w_0$, $w_1$ and $w_2$. 
By setting $w_1=0$ we obtain a simplified version of parameterization (1), as

\begin{equation}\label{padesimp}
w_{\rm de}(a)=\frac{w_0}{1+w_{2}(1-a)}\;.
\end{equation}
In order to desist from each singularity in the wide range of redshifts it is necessary to choose the value of
$w_{2}$ in the interval $-1<w_2<0$. 

As the procedure we follow for parameterization (1), we can obtain 
the DE density and dimensionless Hubble parameter for parameterization (2) as follows:
\begin{eqnarray}
\rho_{\rm de}=\rho^{(0)}_{\rm de} a^{-3(\frac{1+w_0+ w_2}{1+w_2})} [1+w_2 (1-a)]^{-3(\frac{- w_0 w_2}{w_2 (1+w_2)})}\;,\label{rho-simpade1}
\end{eqnarray}

\begin{eqnarray}
E^{2}=\Omega_{\rm r0} a^{-4}+\Omega_{\rm m0} a^{-3} + (1-[\Omega_{\rm r0}+\Omega_{\rm m0}]) \times \nonumber \\ a^{-3(\frac{1+w_0+ w_2}{1+w_2})} \times(1+w_2 - a w_2)^{-3(\frac{ -w_0 w_2}{w_2(1+w_2)})} \;.\label{Esimpade1}
\end{eqnarray}

In Table (\ref{tab:bestfit}) we sum up the chosen values for free parameters characterizing each of models under study. Previously in \citep{Rezaei:2017yyj} we put constraints on the free parameters of these DE parameterizations using different observational data. These values are in the $1\sigma$ confidence level of the best fit parameters, we obtained for the different DE parameterizations using the observational data. These chosen values also are in the $1\sigma$ confidence region of the best fit parameters, which obtained by \citep{Wei:2013jya} using different observational data sets. 

\begin{table}
	\centering
	
	\caption{Values of free parameters for DE parameterizations.
	}
	\begin{tabular}{c  c  c c }
		\hline \hline
		Model  & $parameterization(1)$ & $parameterization(2)$ &$\Lambda$CDM\\
		\hline 
		$\Omega_{\rm m}^{(0)}$ & $0.29$ & $0.27$ & $0.29$\\
		\hline
		$ h $& $0.68$&$0.68$ &$0.68$\\
		\hline
		$ w_0 $ & $-0.83$&$-0.83$ & $-$\\
		\hline
		$ w_1 $ & $-0.09$&$-$ & $-$\\
		\hline
		$ w_2 $ & $-0.68$&$-0.39$ & $-$\\
		\hline \hline
	\end{tabular}\label{tab:bestfit}
\end{table}

\begin{figure} 
	\centering
	\includegraphics[width=0.5\textwidth]{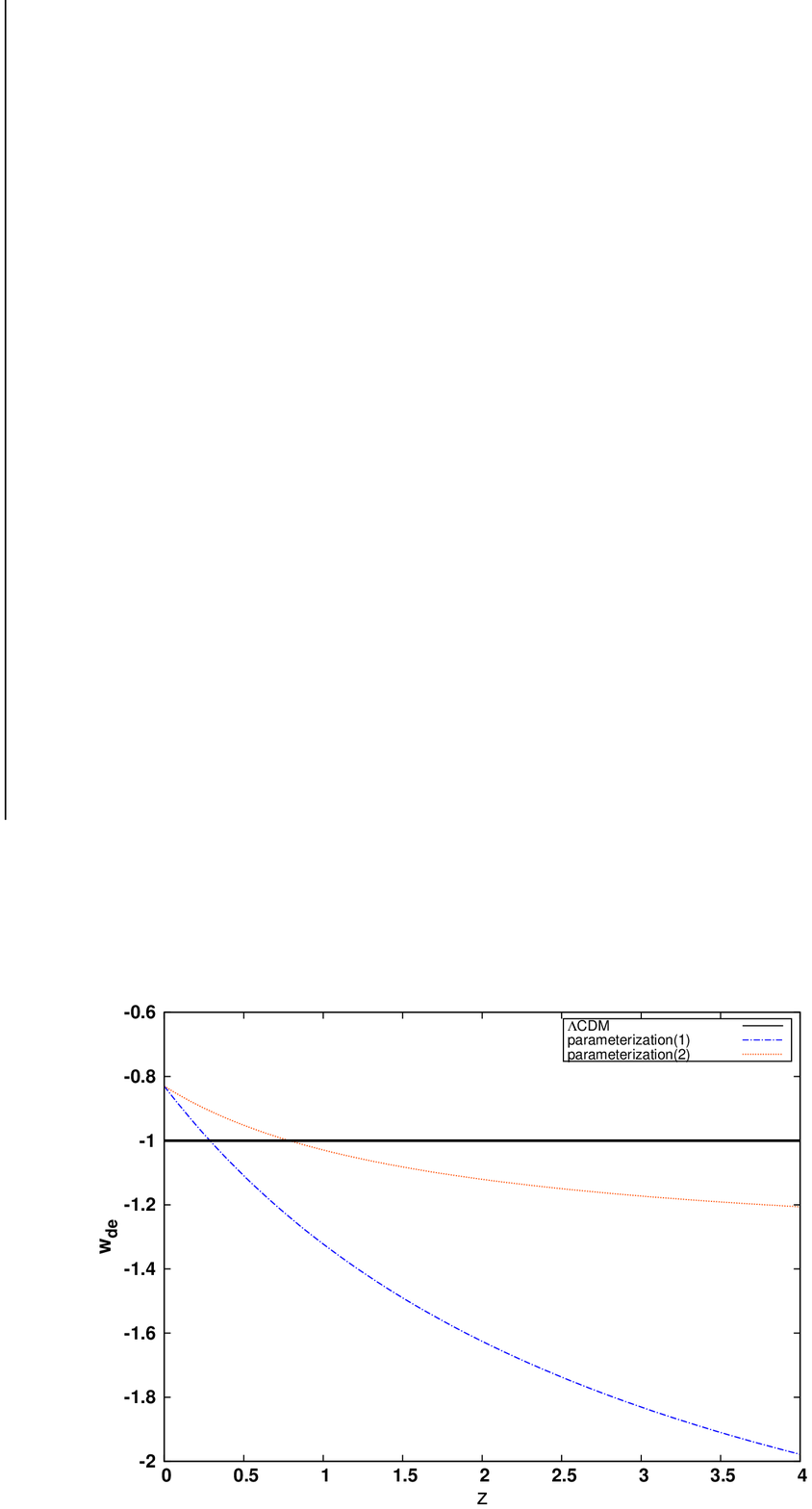}
	\includegraphics[width=0.5\textwidth]{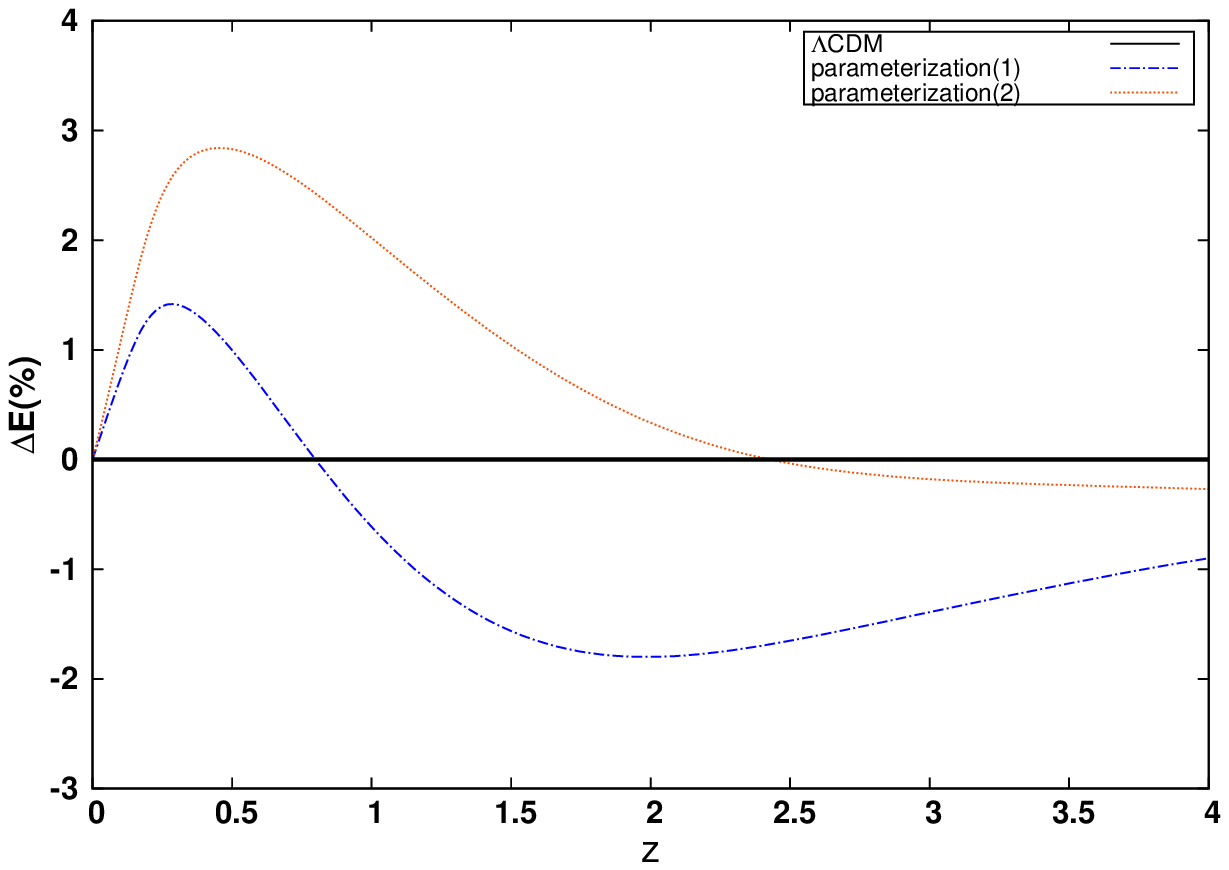}
	\includegraphics[width=0.5\textwidth]{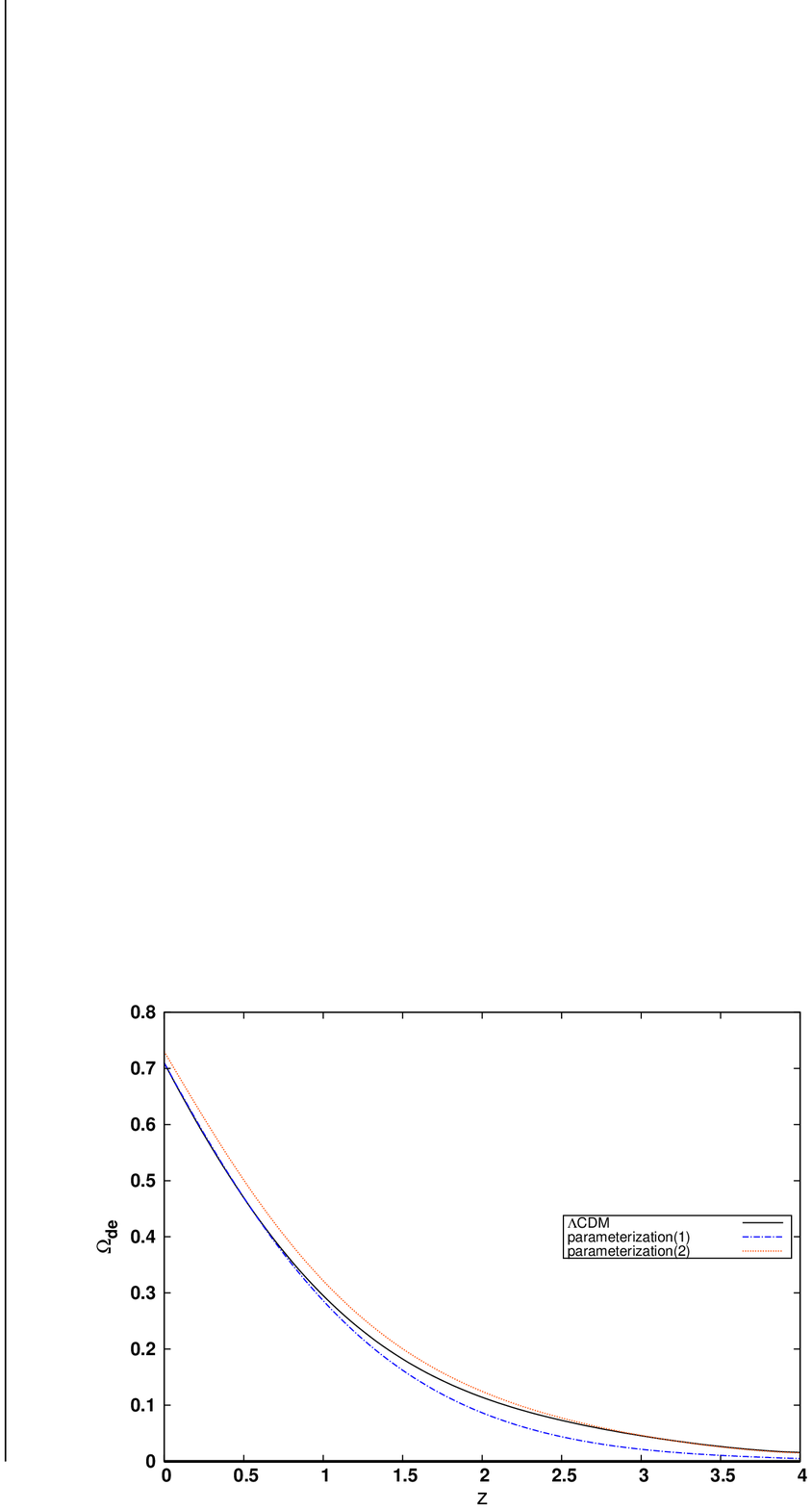}
	\caption{ Top panel:The redshift evolution of the equation of
state parameter of DE parameterizations $w_{\rm d}(z)$, middle panel: the  ratio of dimensionless Hubble parameter of DE parameterizations to the $\Lambda$CDM model and bottom panel: DE density parameter $\Omega_{\rm d}(z)$ for different DE parameterizations. The blue dashed, red dotted and black solid curves correspond to parameterization (1), parameterization (2) and $\Lambda$CDM model respectively. }
	\label{fig:back}
\end{figure}

In the upper panel Fig. (\ref{fig:back}) we present the evolution of $w_{\rm de}$ and in middle panel we plot $\Delta E(\%)=[(E-E_{\rm \Lambda})/E_{\rm \Lambda}]\times 100$ and in bottom panel we show the evolution of $\Omega_{\rm de}$ for our parameterizations using the values of free parameters  presented in Table (\ref{tab:bestfit}). Also, in all panels the reference $\Lambda$CDM model is presented for comparison.

One can see that  at high redshifts the EoS parameter of DE parameterizations considered in this work evolves in phantom regime ($w_{\rm de}<-1$), while at relatively low redshifts it enters in the quintessence region ($-1<w_{\rm de}<-1/3$). From the middle panel of Fig. (\ref{fig:back}), we find that the relative difference
$\Delta E$ for parameterization (1) varies among $\sim-2.0\%$ and $\sim1.4\%$, while in the case of parameterization (2) we have $-0.2 \%\lesssim \Delta E \lesssim 2.9\%$. For both of parameterizations, we have $\Delta E > 0$  at low redshifts ($z \sim 0.5$). In the other meaning in cosmology of PADE parameterizations at low redshifts ($z \sim 0.5$), the rate of the expansion of universe is greater than $\Lambda$ cosmology ones. Lastly, in the bottom panel of Fig.(\ref{fig:back}) the evolution of $\Omega_{\rm de}$ for DE parameterizations and $\Lambda$CDM model are presented. We observe that in all of models, $\Omega_{\rm de}$ tends to zero at high redshifts. At low redshifts parameterization (1) and $\Lambda$CDM have the same value of $\Omega_{\rm de}$, while parameterization (2) has greater value of $\Omega_{\rm de}$. Moving to high redshifts one can see that the value of $\Omega_{\rm de}$ for parameterization (1) becomes smaller than other models. This means that in the case of parameterization (1), DE reaches to its main role in the evolution of the universe at higher redshifts. This result, also can be obtained from the middle panel of Fig. (\ref{fig:back}), where we see that parameterization (2) experiences greater value of Hubble parameter at lower redshifts compared to parameterization (1).

\section{SPHERICAL COLLAPSE IN PADE PARAMETERIZATIONS }\label{growth}

In the current section we investigate the SCM in the framework of PADE parameterization cosmologies. Firstly, we introduce the main equations which are using to obtain the basic parameters of the spherical collapse
model. In order to find the differential equations which can display the evolution of perturbations in both of matter and DE components, several efforts have been done  in the scenario of structure formation. Some of these  efforts have been done to study the evolution of perturbations at high redshifts (in matter dominated universe) \citep{Bernardeau1994,Padmanabhan1996}. A generalization to the universe containing dynamical DE component was done in the work of \citep{Abramo2007}. In the case of inhomogeneous DE in which, DE component can cluster, the fully perturbed equations for the evolution of matter and DE perturbations ($\delta_{\rm m}$ and $\delta_{\rm d}$)
in the non-linear regime are given by\citep{Pace:2014taa,Malekjani:2016edh,Rezaei:2017hon} 

\begin{equation}\label{nl1}
\acute{\delta_{\rm m}}+(1+\delta_{\rm m})\dfrac{\tilde{\theta}}{a}=0\;
\end{equation}
\begin{equation}\label{nl2}
\acute{\delta_{\rm d}}+\dfrac{3(c^2_{\rm eff}-w_{\rm d})}{a}\delta_{\rm d}+(1+w_{\rm d}+(c^2_{\rm eff}+1)\delta_{\rm d})\dfrac{\tilde{\theta}}{a}=0\;
\end{equation}
\begin{eqnarray} \label{nl3}
\tilde{\theta}^{\prime}+(\frac{2}{a}+\frac{E^{\prime}}{E})\tilde{\theta}+\frac{{\tilde{\theta}}^2}{3a}+\frac{3(\Omega_{\rm m}\delta_{\rm m}+(3c^2_{\rm eff}+1)\Omega_{\rm d}\delta_{\rm d})}{2a}=0\;.
\end{eqnarray}
where $\tilde{\theta}=\dfrac{\theta}{H}$ is the dimensionless divergence of the comoving
peculiar velocity for both non-relativistic matter and DE.
Neglecting higher order of $\delta_{\rm m}$ and $\delta_{\rm d}$, one can obtain the linearized Equations. (\ref{nl1},\ref{nl2},\ref{nl3}) as

\begin{equation}\label{l1}
\acute{\delta_{\rm m}}+\dfrac{\tilde{\theta}}{a}=0\;
\end{equation}
\begin{equation}\label{l2}
\acute{\delta_{\rm d}}+\dfrac{3(c^2_{\rm eff}-w_{\rm d})}{a}\delta_{\rm d}+(1+w_{\rm d})\dfrac{\tilde{\theta}}{a}=0\;
\end{equation}
\begin{equation}\label{l3}
\acute{\tilde{\theta}}+(\dfrac{2}{a}+\dfrac{E^{\prime}}{E})\tilde{\theta}+\frac{3(\Omega_{\rm m}\delta_{\rm m}+(3c^2_{\rm eff}+1)\Omega_{\rm d}\delta_{\rm d})}{2a}=0\;.
\end{equation}

For any appropriate initial condition, we can find the redshift evolution of linear overdensities  for matter $(\delta_{\rm m})$ and DE $(\delta_{\rm d})$  by solving Equations. (\ref{l1},\ref{l2},\ref{l3}). If suitable initial conditions were selected, these equations can be used to determine the time evolution of the growth factor. To obtain the suitable initial conditions, we are following constant-infinity method by considering non-linear Eqs.(\ref{nl1},\ref{nl2} \& \ref{nl3})\citep[for detailed discussion, see][]{Herrera:2017epn,Pace:2017qxv}. In the spherical collapse scenario at collapse
redshift $(z_{\rm c})$, the collapsing sphere falls to its center and its non-linear overdensity $\delta_{\rm m}$
becomes formally infinite. Therefore, we should search for a suitable initial
value $(\delta_{\rm mi})$ such that the $\delta_{\rm m}(z_{\rm c})$ obtained from solving the non-linear equations diverges at the chosen collapse redshift. It is clear that our results depend on the selected value of the numerical infinity and on the initial scale factor $a_{\rm i}$ at which we start solving the differential equations. Numerically, we set $a_{\rm i}=10^{-5}$ and the value of numerical
infinity $\delta_{\rm m}(z_{\rm c})$ to be of order ${10}^8$ in order to provide conditions which described in \citep{Pace:2017qxv}. Once $\delta_{\rm mi}$ is found, we apply this value as one of the initial conditions we need to solve the linear differential equations (\ref{l1},\ref{l2}\&\ref{l3}) to obtain one of the main parameters in SCM scenario, the linear threshold parameter $\delta_{\rm c}$. In fact in the context of SCM when $\delta_{\rm m}^{linear}\geq\delta_{\rm c}$ the corresponding perturbed region is virialized. Since we want to solve three differential equations, except  $\delta_{\rm mi}$  we need two other initial
conditions. These two remained quantities are the initial
values of $\delta_{\rm di}$, the DE overdensity  and ${\tilde{\theta}}_{\rm i}$, the peculiar velocity perturbation, which both of them can be evaluated using $\delta_{\rm mi}$, using these equations \citep{Batista:2013oca,Pace:2014taa,Rezaei:2017hon}:
\begin{equation}\label{in1}
\delta_{\rm di}=\dfrac{n}{n-3w_{\rm d}}(1+w_{\rm di})\delta_{\rm mi}\;
\end{equation}
\begin{equation}\label{in2}
{\tilde{\theta}}_{\rm i}=-n\delta_{\rm mi}\;.
\end{equation}

In the Einstein de-Sitter (EdS) universe as a special case, we have $n=1$. However, in the other DE cosmologies it has been shown that $n$ has a small deviation from unity \citep{Batista:2013oca}. Since at high redshifts the contribution of DE is negligible, we approximately set $n=1$ in equations (\ref{in1} \& \ref{in2}) to calculate two other initial conditions in order to solve linear equations (\ref{l1},\ref{l2} \& \ref{l3}).
In order to investigate SCM we can follow two approaches: in the first approach the DE component is homogeneous ($\delta_{\rm de}\equiv 0$) and only dark matter and baryons are allowed to cluster ($\delta_{\rm m}\ne 0$). In the second approach, both of matter and DE components  are allowed to cluster.
Clustering of dark energy as a more general case is defined based on the value of effective sound speed ${c^2_{\rm eff}}\equiv\delta P_{\rm d}/\delta \rho_{\rm d}c^2$. In the case of DE with ${c^2_{\rm eff}}\sim1$, we have only slight DE perturbations at highly nonlinear phase of evolution of dark matter halo. The amplitudes of relevant density and velocity perturbations of such dark energy at turnaround phase of halo are of order $10^{-6}$ and $10^{-4}$ accordingly \citep{Novosyadlyj:2016lyc}. Hence DE perturbations will reach of own turnaround point in far future. Therefore in comparison with dark matter perturbations we can ignore $\delta_{\rm d}$ when we solve the systems of equations(\ref{nl1},\ref{nl2} \& \ref{nl3}) and (\ref{l1},\ref{l2} \& \ref{l3}). In the second scenario, clustered DE, by setting ${c^2_{\rm eff}}=0$ DE fluctuations can 
reach the point of turnaround and collapse together with dark matter and so affect the evolution of $\delta_{\rm m}$ \citep{Creminelli2010,Batista:2013oca,Malekjani:2018qcz,Mehrabi:2018dru}. The density of dark energy in this scenario can be large and essentially can affect the virialization of halo, its total mass , its density profile, mass function and so on. These effects can be useful for discrimination of different dark energy models and to distinguish among DE and cosmological constant.

\subsection{growth factor}
Here we follow the linear growth of perturbations of non-relativistic
dust matter by solving coupled linear equations (\ref{l1},\ref{l2} \& \ref{l3}). One can compute the linear growth factor as one of the main parameters in spherical collapse scenario by \citep[for similar discussion, see also][]{Copeland2006, Nesseris:2007pa,Tsujikawa2008, Pettorino:2008ez, Lee2011}

\begin{equation}\label{gfactor}
D_+(a) =\delta_{\rm m}(a)/\delta_{\rm m}(a = 1)\;.
\end{equation}

\begin{figure} 
	\centering
	\includegraphics[width=0.5\textwidth]{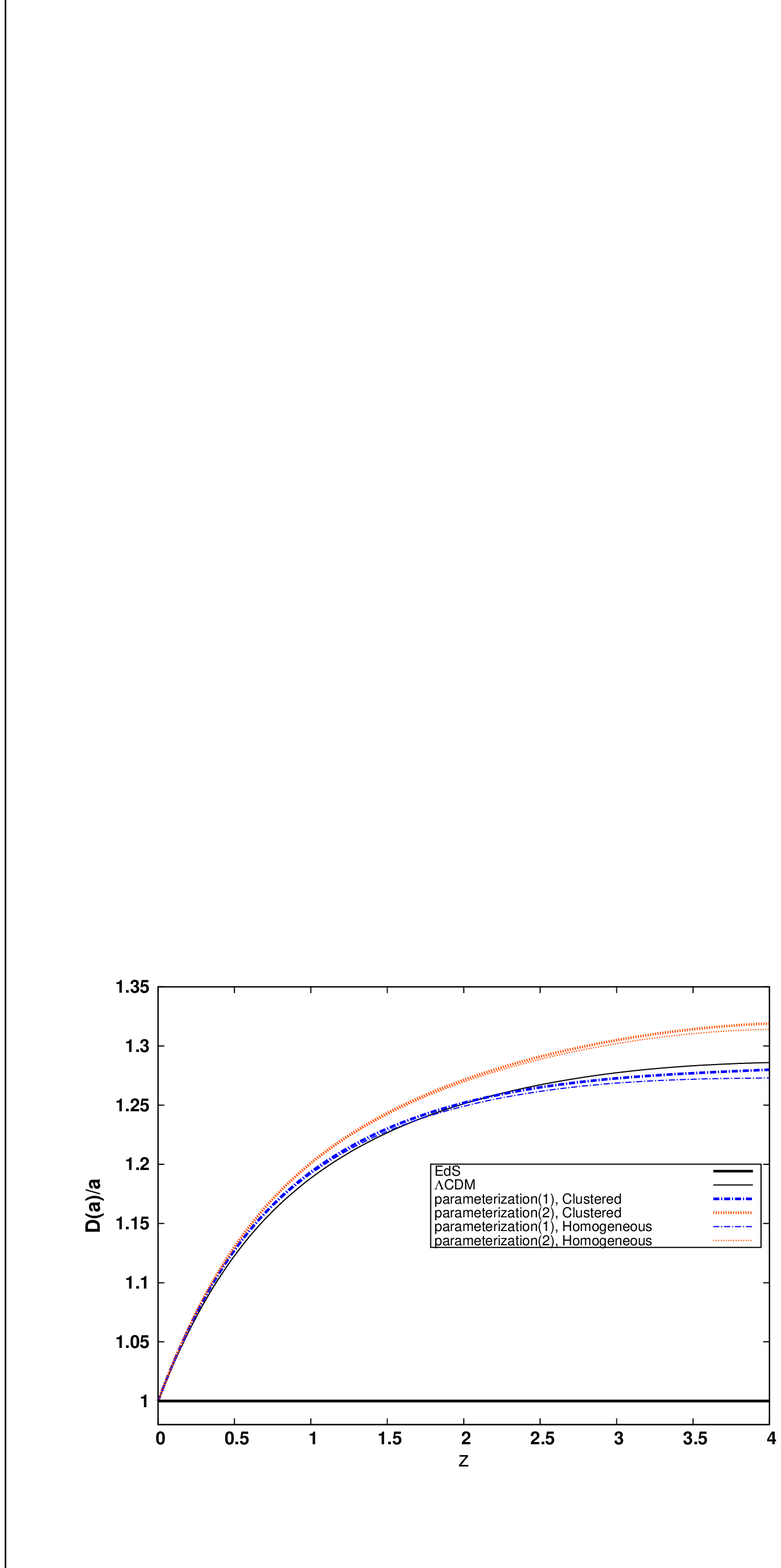}
  	\caption{ The redshift evolution of growth factor for different parameterizations considered in this work. Thick and thin curves represent clustered and homogeneous DE respectively. 
The reference $\Lambda$CDM (EdS) model is shown by thin (thick) solid black line.}
	\label{fig:d}
\end{figure}

Fig.\ref{fig:d} shows the  redshift evolution of the growth factor normalized at $z = 0$ and divided by the scale factor $a$. In the EdS model (thick black line) at any time we have $D_+(a)/a=1$, which  shows that the growth of matter perturbations $\delta_{\rm m}$ is the same at all redshifts. In the case of $\Lambda$ cosmology (thin black line) the growth factor is higher than the EdS model throughout
its history, but falls for lower redshifts, because at late times the cosmological
constant dominates the energy budget of the universe
and suppresses the amplitude of perturbations. On the other hand, a
larger growth factor in the $\Lambda$ cosmology case at higher redshift, shows
that the growth of matter perturbations will be stronger than in an EdS universe at early times. In the case of DE parameterizations considered in this work,  same as $\Lambda$CDM model we  see that the growth factor is higher than the EdS model.  This result is expected since in DE models, DE suppresses the growth of matter perturbations, while in EdS universe this suppression does not exist. Therefore in DE models, the initial matter perturbations should grow with larger values of growth factor than EdS universe to exhibit the large scale structures we observe today. In comparison with $\Lambda$CDM model, parameterization (2) has a
larger growth factor, while parameterization (1) experiences lower values of growth factor at relatively higher redshifts. Since parameterization (2) has a greater value of Hubble parameter at relatively high redshifts [see middle panel of Fig. (\ref{fig:back})] , the growth rate of perturbations in it must be larger than parameterization (1) ones, till both of them reach to same value of structures at present time. Moreover, for both of our parameterizations, the growth factor in clustered DE cases is bigger than those of obtained in homogeneous DE cases respectively.

\subsection{ linear overdensity parameter $\delta_{\rm c}$}

\begin{figure} 
	\centering
	\includegraphics[width=0.5\textwidth]{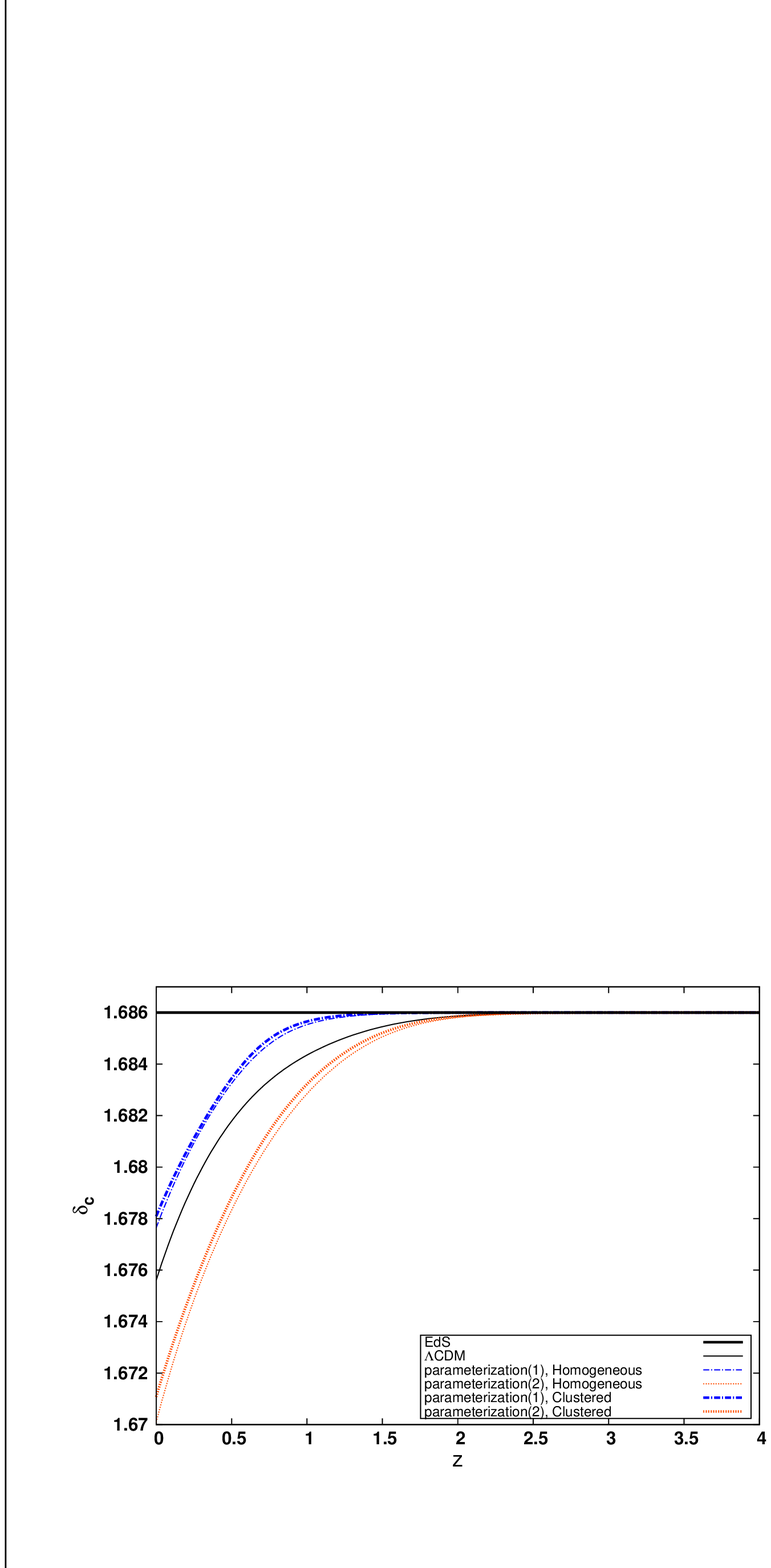}
    \includegraphics[width=0.5\textwidth]{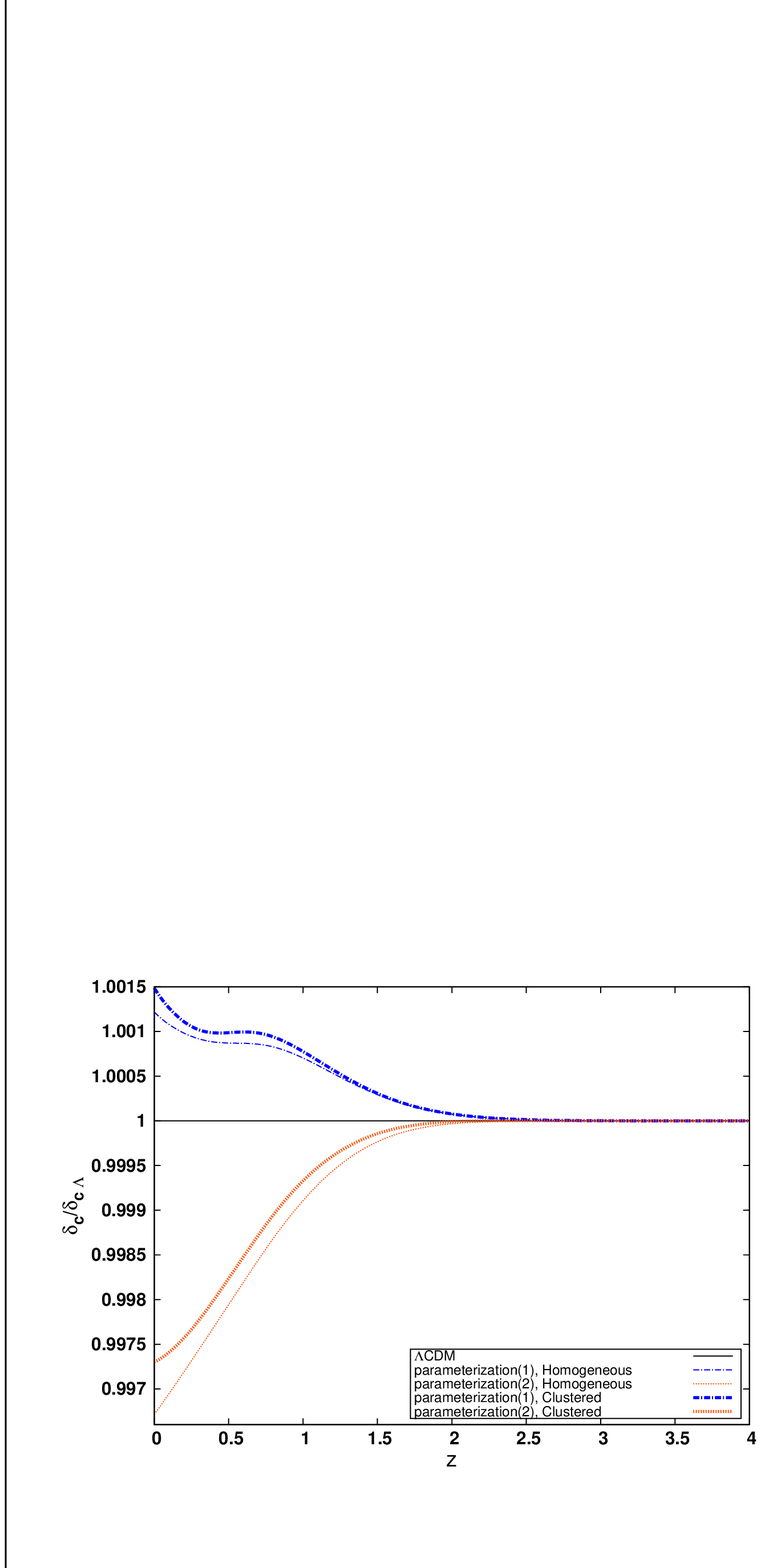}
	\caption{ Top panel: linear overdensity $(\delta_{\rm c})$ parameter
 versus $z_{\rm c}$ and bottom panel: the ratio of parameter $\delta_{\rm c}$
of DE parameterizations to that of the $\Lambda$CDM versus $z_{\rm c}$, for different DE parameterizations considered in this study. Line styles are same as Fig.(\ref{fig:d}).}
	\label{fig:dc}
\end{figure}

Now we calculate one of the  main quantities of SCM, the linear overdensity parameter $\delta_{\rm c}$ in the context of PADE parameterizations for the EoS of DE. This parameter together
with the linear growth factor $D_+(z)$ are used to compute the mass function of virialized halos \citep[see e.g.][]{Press1974,Sheth:1999su,Sheth2002}. Our results for the evolution of $\delta_{\rm c}$ are presented in Fig.(\ref{fig:dc}). In the top panel we plot the redshift evolution of the linear overdensity parameter $\delta_{\rm c}$  and in the bottom panel we plot the ratio of the linear overdensity parameter for different DE parameterizations to that of $\Lambda$CDM.

In Fig.(\ref{fig:dc}) we observe that parameterization (1) (parameterization (2)) always have a higher (lower) $\delta_{\rm c}(z)$ with respect to the $\Lambda$CDM model. We also see that the value of $\delta_{\rm c}$ in clustered DE parameterizations is larger compare to homogeneous ones. This result indicates that in these parameterizations, clustering of DE can support the formation of structures. This result is in good agreement with the results of \citep{Abramo:2007iu} in which authors indicated that fluctuations in phantom DE components enhance the growth of matter perturbations. The reason is that in phantom regime dark matter overdensities lead to voids (underdensities) in DE component. Because of gravitationally repulsive nature of DE, underdensities in it help the matter perturbations to grow faster \citep{Abramo:2007iu}. The latter results, were obtained for DE parameterizations that are phantom at all times. While we obtain same results for DE parameterizations which are in phantom regime at a wide range of redshifts and became non-phantom just at very low redshifts. The difference between  $\delta_{\rm c}$ of parameterization (1) (parameterization (2)) compared to that of  $\Lambda$CDM is smaller than $0.15 \%$ ($0.4 \%$). Recent PADE parameterizations similar to  $\Lambda$CDM model, asymptotically approach the EdS limit at relatively high redshifts, where  the effects of DE can be ignored.

\subsection{virial overdensity parameter $\Delta_{\rm vir}$}

The other important parameter in SCM is the virial overdensity parameter $\Delta_{\rm vir}$. This parameter is applied to obtain the size of dark matter halos. This parameter has the form $\Delta_{\rm vir} = \delta_{\rm nl} + 1 = \zeta(x/y)^3$, where we have $x = a/a_{\rm t}$ as the normalized scale factor $a$ and $y$ as the radius of collapsing sphere normalized to its value at the turn-around redshift \citep{Wang1998}. Also $\zeta$ is the ratio of matter density inside the overdense sphere to its value at out of sphere (at background) at turn-around redshift \citep{Naderi2015}. In order to obtain $a_{\rm t}$ we can solve set of non-linear equations (\ref{nl1},\ref{nl2} \& \ref{nl3}) and find the value of $\log (\delta_{\rm nl} + 1)/a^3$. This value is the inverse of the radius of overdense sphere and its minimum denotes the maximum radius which take place at turn around scale factor. Moreover, to determine the value of  $\zeta$ we can compute $\delta_{\rm nl} + 1$ by integrating Eq.\ref{nl1} up to turn around scale factor \citep[for more details see][]{Pace2010,Naderi2015}.

\begin{figure} 
	\centering
	\includegraphics[width=0.5\textwidth]{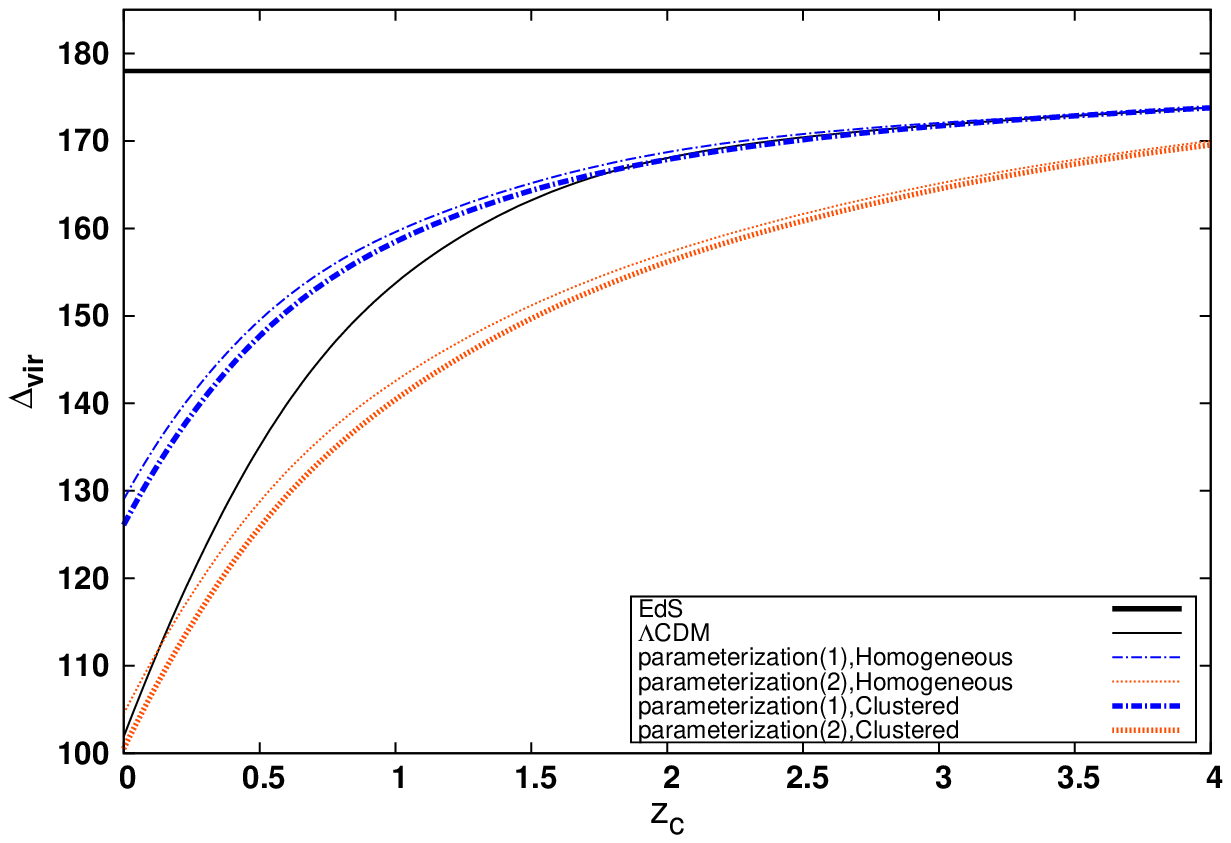}
     \includegraphics[width=0.5\textwidth]{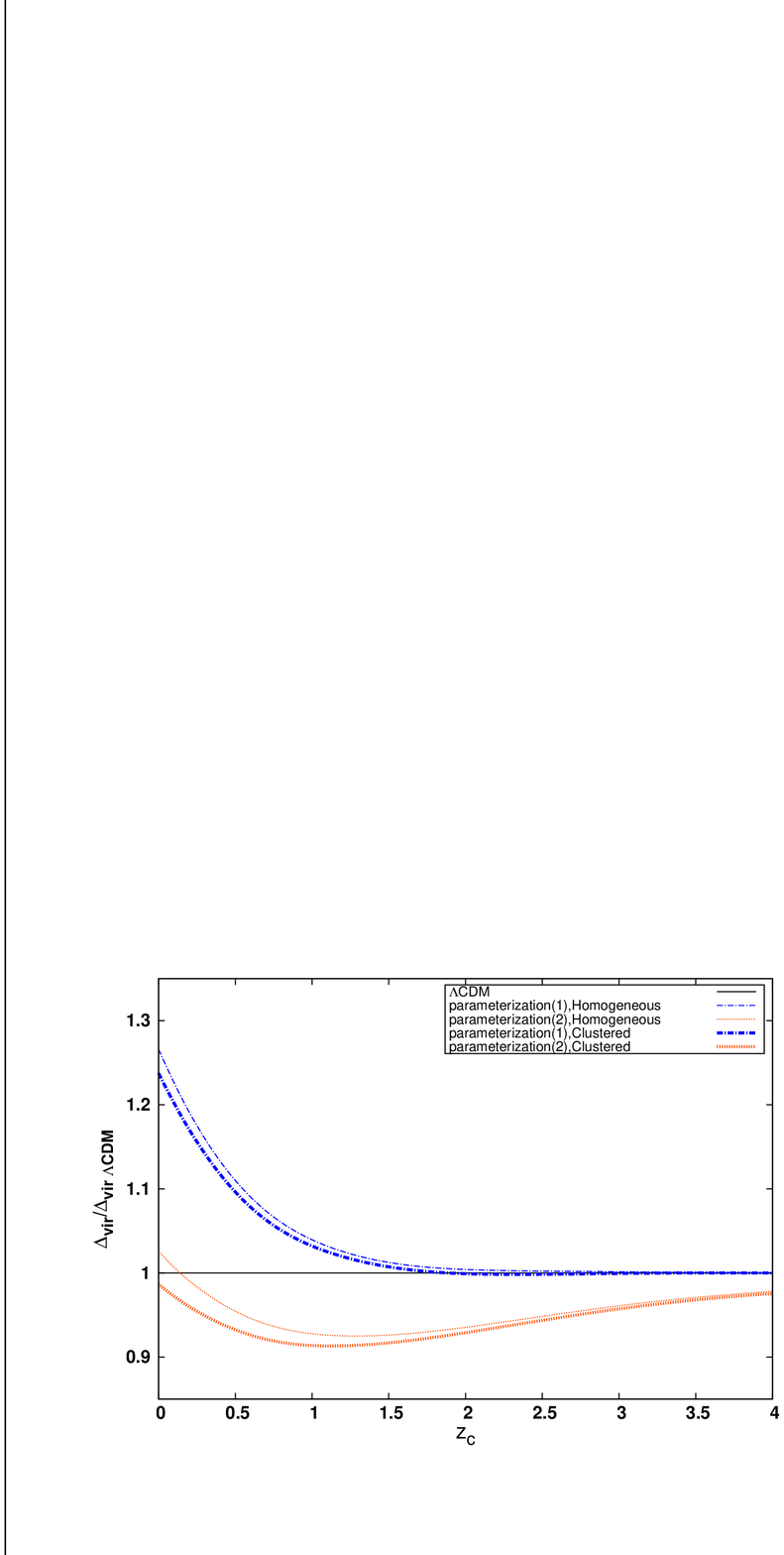}
    \includegraphics[width=0.5\textwidth]{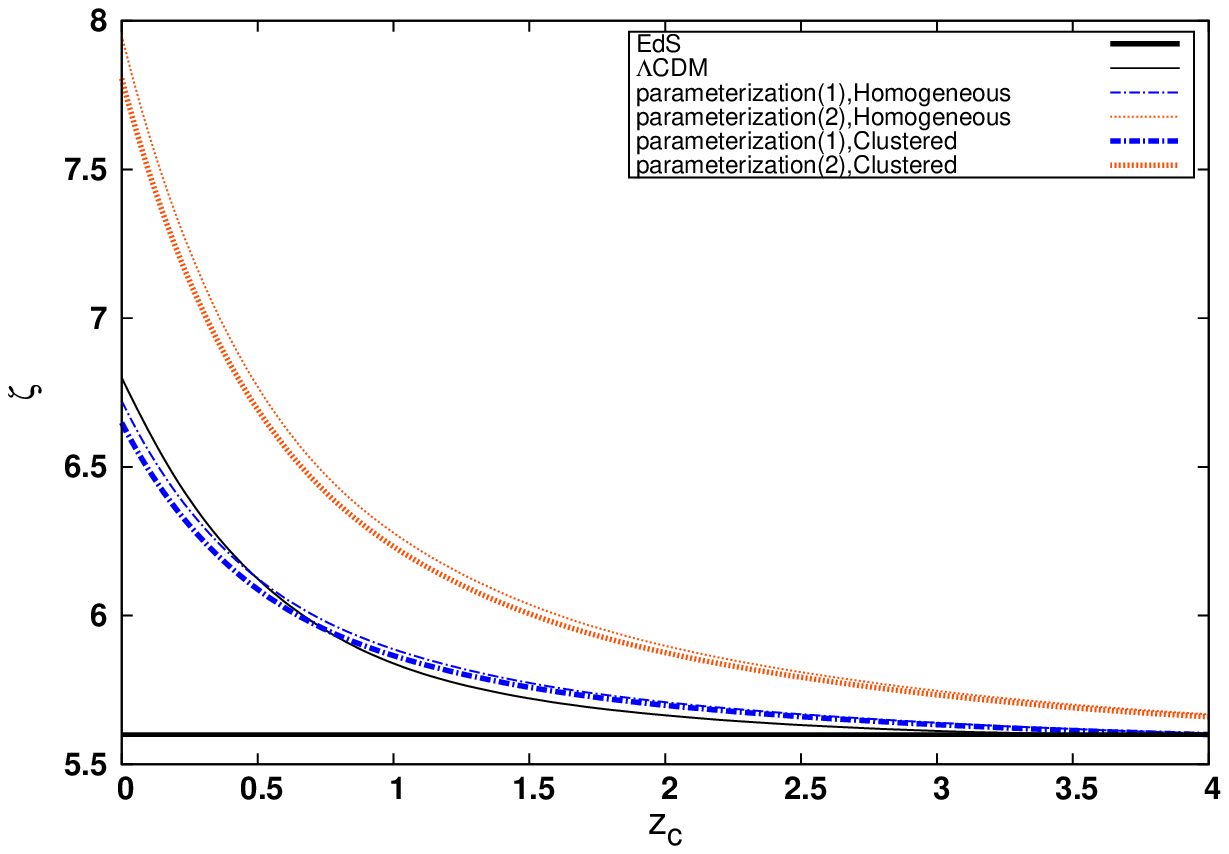}
	\caption{Top panel: the evolution of virial overdensity parameter ($\Delta_{\rm vir}(z)$) versus  $z_{\rm c}$, middle panel: normalized $\Delta_{\rm vir}(z)$ to $\Lambda$CDM value and  bottom panel: turn-around overdensity parameter ($\zeta$) for different DE parameterizations. Line styles are same as Fig.(\ref{fig:d}).} 
\label{fig:dv}
\end{figure}

In the  top panel of Fig.\ref{fig:dv} we present the redshift evolution of the virial overdensity
parameter $\Delta_{\rm vir}(z)$  and in the middle panel we plot the value of  $\Delta_{\rm vir}(z)$ normalized to  $\Delta_{\rm vir}(z)$ for $\Lambda$ cosmology. In the both of DE parameterizations same as $\Lambda$CDM model, $\Delta_{\rm vir}$ tends to $178$, the value of $\Delta_{\rm vir}$ for EdS model, at relatively higher redshifts. This result is expected, because at high redshifts the role of DE in the evolution of the universe become insignificant and thus results of different DE models tends to those of Eds model. At low redshifts the decrements of $\Delta_{\rm vir}$ shows that in cosmologies with DE or $\Lambda$ component, low dense virialized halos are formed compare to EdS model. This result is expected, because DE component oppose the collapsing of halos. Between parameterizations under study in this work, in the case of parameterization (1)  the density of matter component in virialized halos is about $25\%$  higher than that of $\Lambda$CDM model, while in parameterization (2) this difference reduces to $ \leq \pm2\%$ at $z_c=0$. More ever one can observe in both of parameterizations that $\Delta_{\rm vir}$ in homogeneous DE scenarios is larger than clustered DE scenarios. Finally,  one can see the evolution of $\zeta$, turn-around overdensity parameter in the bottom panel of Fig.\ref{fig:dv}. As expected, at high redshifts, $\zeta$ tends to the EdS value $\zeta=5.6$, because at these redshifts the matter component dominated the energy budget of universe.   In both clustered and homogeneous scenarios of parameterization (2), the value of $\zeta$ is larger than those of parameterization (1) and $\Lambda$CDM model. These results can be concluded from the middle panel of Fig.\ref{fig:back}. In this figure we saw that the value of the expansion rate of the universe in parameterization (2) is greater than those of parameterization (1) and $\Lambda$CDM. Therefore, in the case of  parameterization (2) the overdense sphere decouples from cosmic flow at relatively higher redshifts (or equally at higher value of overdensity). Moreover, in both of parameterizations, value of $\zeta$ is smaller for clustered DE cases compared to homogeneous cases. This result shows that in homogeneous DE scenarios, the perturbed spherical region separates from its background with higher value of overdensity. This result confirms the consequences which we saw in upper panels of Fig.\ref{fig:dv}, where the relatively high dense virialized halos were forming in the homogeneous versions of DE parameterizations.

\section{MASS FUNCTION AND ABUNDANCE OF VIRIALIZED DM HALOS}\label{sec:mass}

While N-body simulation is an important tool for investigation of structure formation in cosmology, it is very
time consuming. Therefore semi-analytic methods can be useful alternatives. In the recent section using the well known  Press-Schechter formalism as a semi-analytic method, we compute the number counts of cluster-size halos 
in DE cosmologies. In Press-Schechter formalism we can express the plenty of
virialized halos of dark matter in terms of their mass
\citep{Press1974}. The value of comoving number density of virialized
objects with masses from $M$ to $M+dM$ at redshift $z$ obtained as follows \citep{Press1974,Bond1991}:
\begin{equation}\label{press}
\dfrac{dn(M,z)}{dM}=\dfrac{\rho_{\rm m0}}{M}\dfrac{d\sigma^{-1}}{dM}f(\nu)\;
\end{equation}

where $\rho_{\rm m0}$ is the value of matter density in background at $z=0$, $\nu(M,z)=\delta_{\rm c} / \sigma$,  $\sigma$ is the r.m.s. of the mass fluctuations in a sphere region which contain mass $M$.

 Although, the standard form of  Press-Schechter mass function with $f(\nu)=\sqrt{{2}/{\pi}} \nu e^{-\frac{\nu}{2}}$ which discussed in \citep{Press1974,Bond1991} can provide
a good approximation of the predicted number density of halos, it fails
by predicting approximation too many low-mass halos  and too few high-mass ones \citep{Sheth1999,Sheth2002,Lima:2004np}. Thus, in this study we apply another well known fitting formula which first proposed in \cite{Sheth1999}:
\begin{equation}\label{sheth}
f(\nu)=0.2709\sqrt{\dfrac{2}{\pi}}(1+1.1096\nu^{0.6})exp(-\dfrac{0.707 \nu^2}{2})\;
\end{equation}
In a Gaussian density field, $\sigma$ is given by:
\begin{equation}\label{sigma}
\sigma^2(R)=\dfrac{1}{2 \pi^2}{\int_0}^\infty k^2 P(k) W^2(kR) dk\;
\end{equation}
where $R=(3M/4\pi \rho_{\rm m0})^{1/3}$ is the radius of the spherical overdense region, $W(kR)$ is the Fourier transform of a spherical top-hat profile with radius $R$ and $P(k)$ is
the linear power spectrum of density fluctuations \citep{Peebles1993}. To obtain the value of $\sigma$, we follow the procedure presented in \citep{Abramo2007}. Following on \cite{Ade:2015xua}, we use the normalization of matter power spectrum $\sigma_8=0.815$ for $\Lambda$CDM cosmology. The number density of virialized halos above a certain value of mass $M$ at $z_c$, the collapse redshift obtained by

\begin{equation}\label{nn}
N(\> M,z)={\int_0}^\infty \dfrac{dn(z)}{dM'}dM'\;.
\end{equation}
The above limit of integration in Eq.\ref{nn} is $M=10^{18}M_{sun}h^{-1}$ which such gigantic structures could not in practice be observed.
Now we can calculate the number density of virialized
halos in both homogeneous and clustered DE scenarios using equations(\ref{press} \& \ref{nn}). In this way the total mass of a halo is equal to the mass of pressureless matter perturbations. However, the virialisation of
dark matter perturbations in the non-linear regime can not be independent from the
properties of DE \citep{Lahav1991,Maor2005,Creminelli2010,Basse2011}. Thus, in clustered
DE scenarios, we should consider the contribution of perturbated DE components to the total mass of the halos \citep{Creminelli2010,Basse2011,Batista:2013oca}. Based on the behavior of $w_{\rm de}(z)$, DE can reduce or enhance the total mass of the virialized halo.
One can obtain $\epsilon(z)$, the ratio of DE mass to be taken into account with respect to the
mass of dark matter, from:

\begin{equation}\label{epsil}
\epsilon(z)=\dfrac{m_{\rm DE}}{m_{\rm DM}}\;
\end{equation}

where the value of $m_{\rm DE}$ depends on what we consider as the
mass of  DE component. When one only considers the contribution of
the perturbations of DE, the $m_{\rm DE}$ takes the form:

\begin{equation}\label{mdep}
{m_{\rm DE}}^{Perturbed}=4 \pi \bar{\rho}_{\rm DE}{\int_0}^{R_{\rm vir}} dR R^2 \delta_{\rm DE}(1+3{c_{\rm eff}}^2)\;.
\end{equation}
In the other hand, if we assume both DE contributions of
perturbation and background level, the total mass of DE in virialized halos takes this new form:
\begin{equation}\label{mdet}
{m_{\rm DE}}^{Total}=4 \pi \bar{\rho}_{\rm DE}{\int_0}^{R_{\rm vir}} dR R^2 [(1+3 w_{\rm DE})+ \delta_{\rm DE}(1+3{c_{\rm eff}}^2)]\;
\end{equation}
The quantities inside a spherical collapsing region in the framework of the top-hat profile, evolve only with cosmic time. Thus from Eq.(\ref{mdep}) one can find:

\begin{equation}\label{epsil1}
\epsilon(z)=\dfrac{\Omega_{\rm DE}}{\Omega_{\rm DM}}\dfrac{\delta_{\rm DE}}{1+\delta_{\rm DM}}\;
\end{equation}
 and from Eq.(\ref{mdet}) we can obtain:

\begin{equation}\label{epsil2}
\epsilon(z)=\dfrac{\Omega_{\rm DE}}{\Omega_{\rm DM}}\dfrac{1+3 w_{\rm DE}+\delta_{\rm DE}}{1+\delta_{\rm DM}}\;
\end{equation}

\begin{figure} 
	\centering
	\includegraphics[width=8cm]{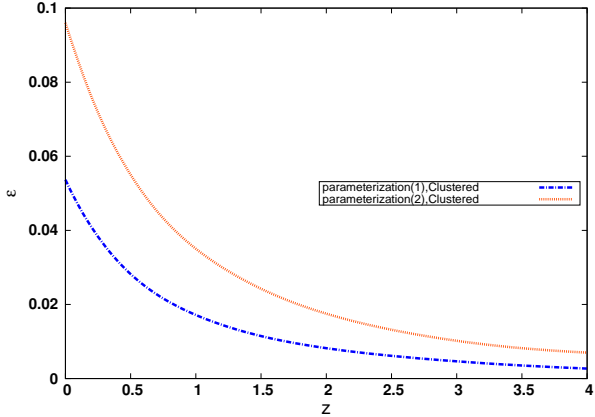}
	\caption{Evolution of $\epsilon(z)$, the mass ratio of DE to DM based on Eq.\ref{epsil1} for different  parameterizations.}
	\label{fig:eps}
\end{figure}

The mass of dark matter also is obtained from \citep[see also][]{Batista:2013oca}:

\begin{equation}\label{mdm}
{m_{\rm DM}}=4 \pi \bar{\rho}_{\rm DM}{\int_0}^{R_{\rm vir}} dR R^2 (1+ \delta_{\rm DM})\;.
\end{equation}
In Fig.(\ref{fig:eps}) we plot the evolution of $\epsilon(z)$ using Eq.(\ref{epsil1}) as the definition of DE mass. We observe that, at high redshift,
where the role of DE is less important, $\epsilon$ for both of parameterizations becomes negligible. This parameter has a greater value in the case of parameterization (2).

To obtain the number density of virialized halos in clustered DE scenario, we should assume DE mass correction. Following the procedure outlined
in \cite{Batista:2013oca,Pace:2014taa}, the mass of virialized halos in clustered DE scenarios is $M(1-\epsilon)$. Hence, the corrected form of mass function can be written as \citep{Batista:2013oca} 
\begin{equation}\label{presscor}
\dfrac{dn(M,z)}{dM}=\dfrac{\rho_{\rm m0}}{M(1-\epsilon)}\dfrac{d \nu(M,z)}{dM}f(\nu)\;.
\end{equation}
In the case of clustered DE models, inserting Eq.(\ref{presscor}) into Eq.(\ref{nn}) we can compute the number density of virialized halos.

\begin{figure*} 
	\centering
	\includegraphics[width=8cm]{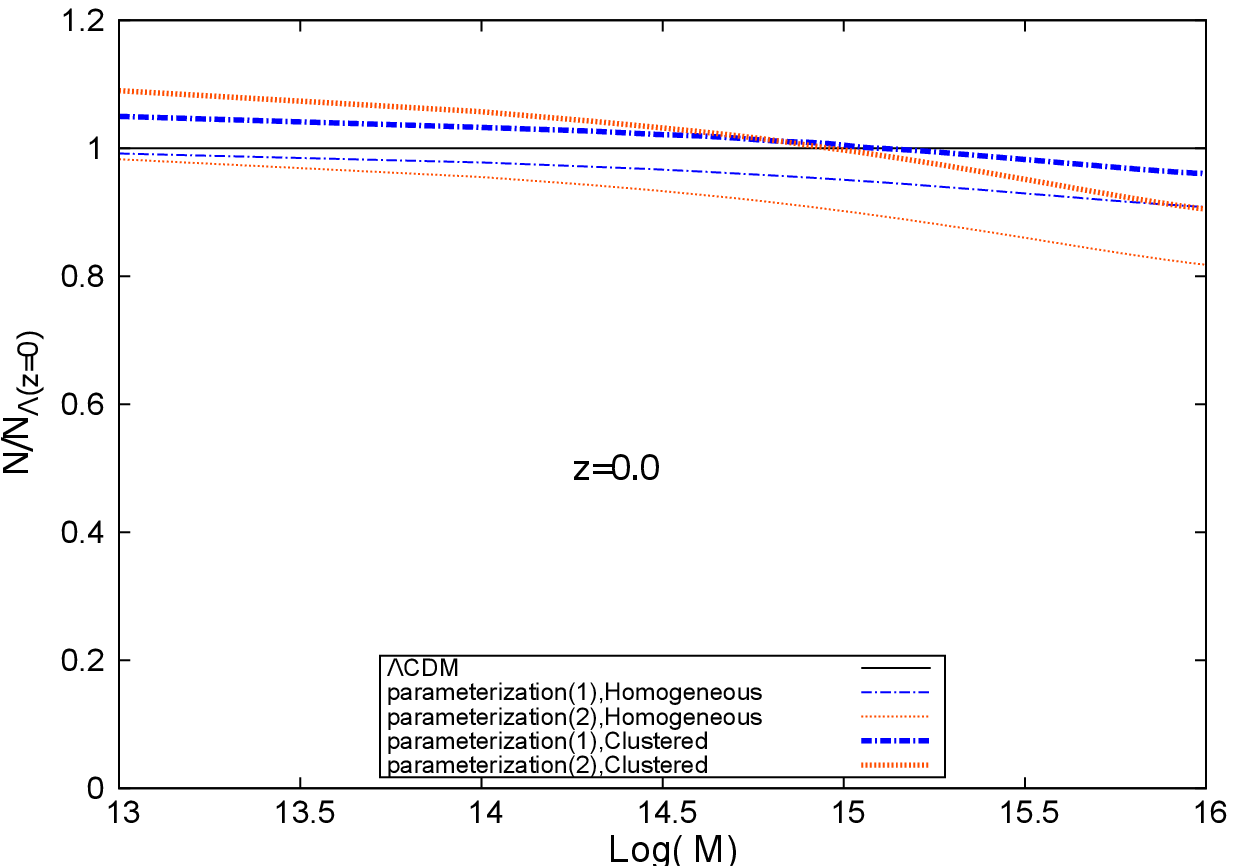}
    \includegraphics[width=8cm]{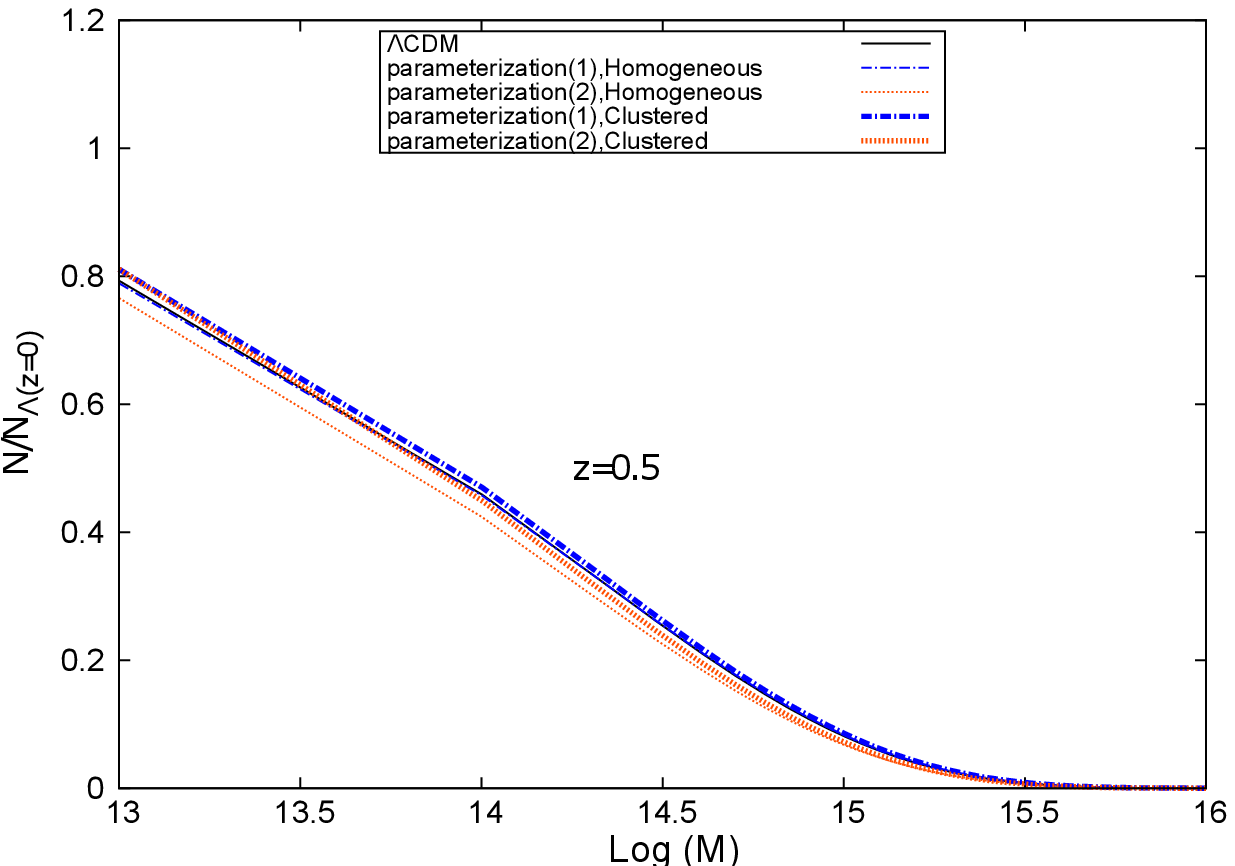}
	\includegraphics[width=8cm]{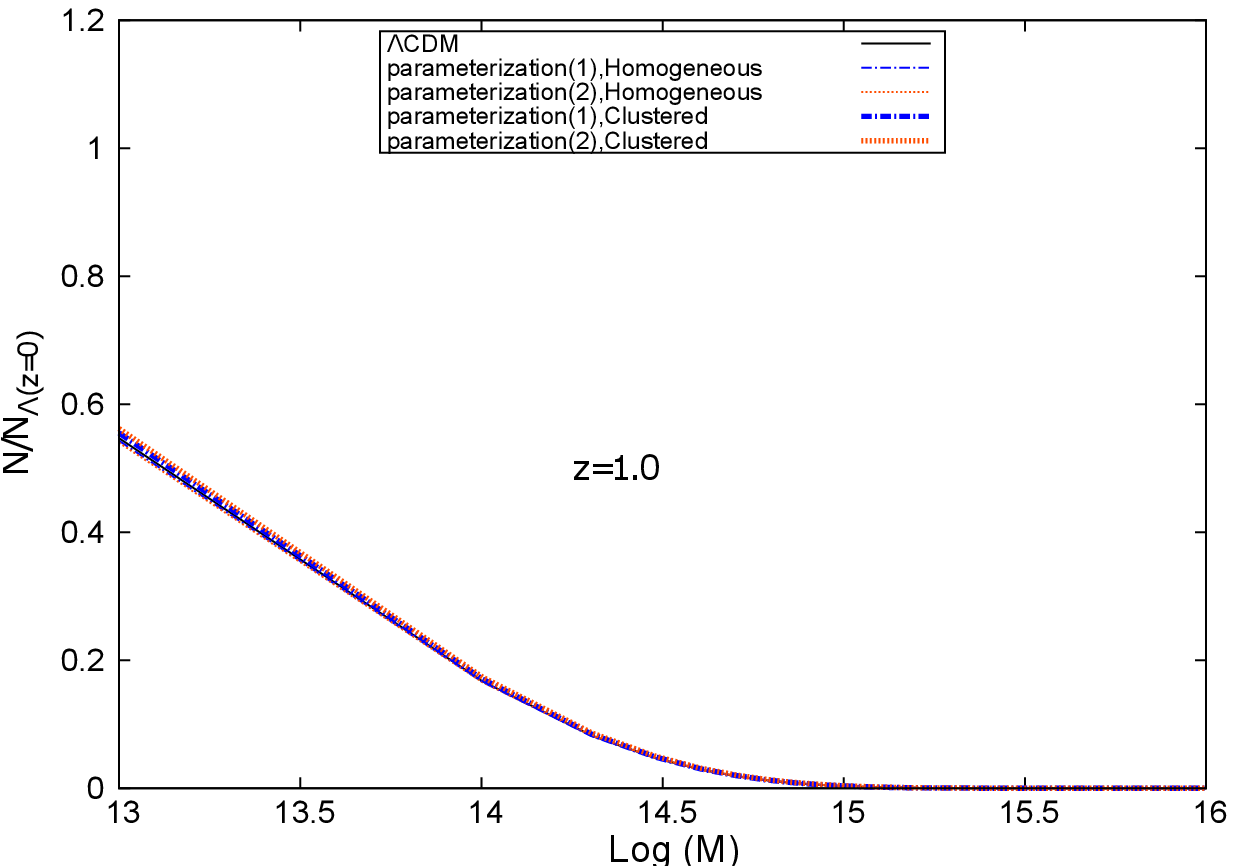}
    \includegraphics[width=8cm]{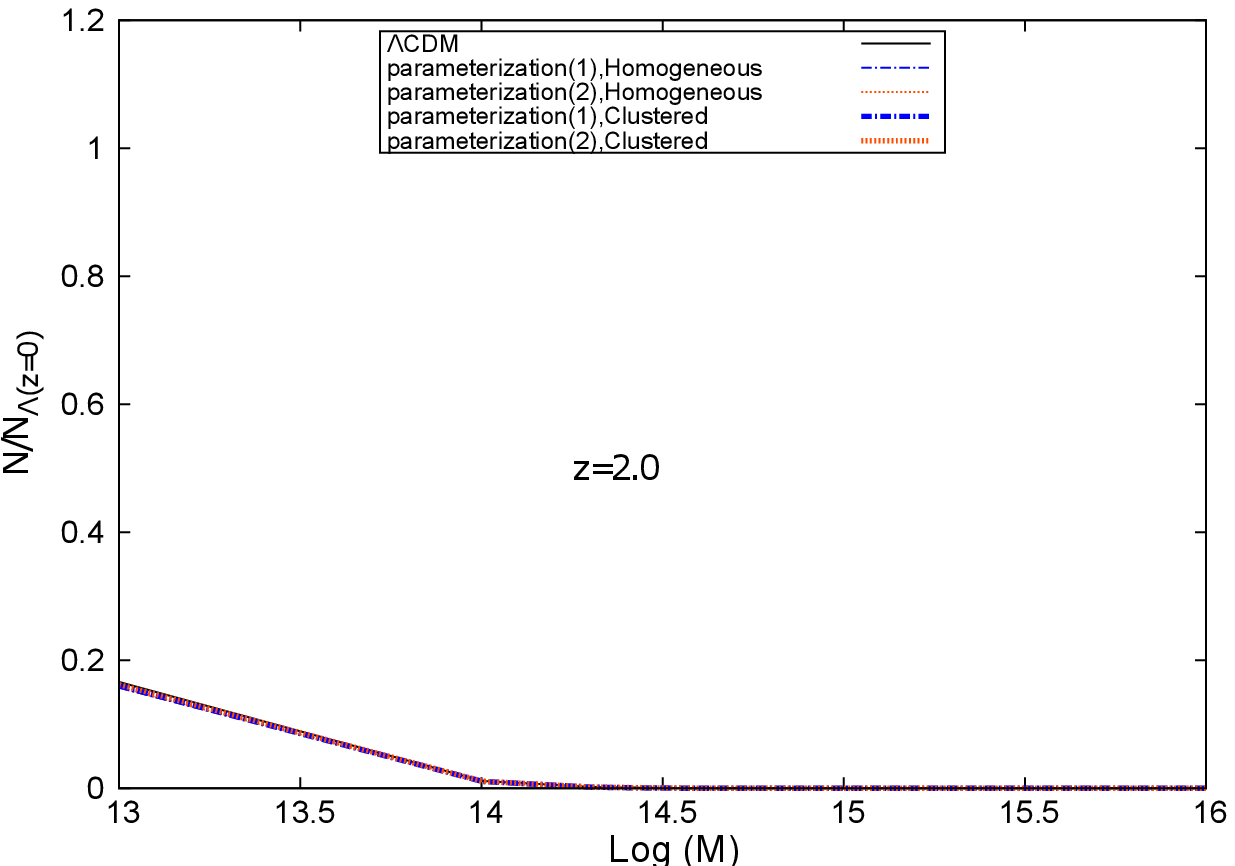}
	
	\caption{Ratio of the number of cluster-size halos above a given mass $M$ for different parameterizations considered in this work to the concordance $\Lambda$CDM model at $z=0$ , $z=0.5$, $z=1.0$ and $z=2.0$.}
	\label{fig:nst}
\end{figure*}

After computing the number density of cluster-size halos at different redshifts: $z=0.0, 0.5, 1.0$ \& $2.0$ for
different  parameterizations considered in this work, we plot the numerical results of our analysis in Fig.(\ref{fig:nst}).  In this way, we normalize the results of our DE parameterizations by that of the $\Lambda$CDM model at $z=0$. We can summarize the main results of this section as follows:
At the present time, $z=0$, we observe that Sheth-Tormen mass function for the both of parameterizations in homogeneous case, predicts less abundance of virialized halos than the $\Lambda$ cosmology at both the low and high mass tails. While, in the clustered DE scenario for both of parameterizations, we have more (less) abundance of halos than the $\Lambda$CDM model for low (high) mass objects. We observe that at this time, the differences between DE parameterizations and concordance $\Lambda$CDM is considerable at all mass scales. In particular, in the case of clusters with mass above $M=10^{13}M_{sun}h^{-1}$, at $z=0$, Sheth-Tormen mass function predicts number density of halos roughly in the clustered (homogeneous) case of parameterization(1) $5\%$ higher ($0.8\%$ lower) than value which predicted for $\Lambda$CDM model. These results for parameterization(2) are $9\%$ for clustered ($-2\%$ for homogeneous) DE scenario. 

\begin{figure*} 
	\centering
	\includegraphics[width=0.87 \textwidth]{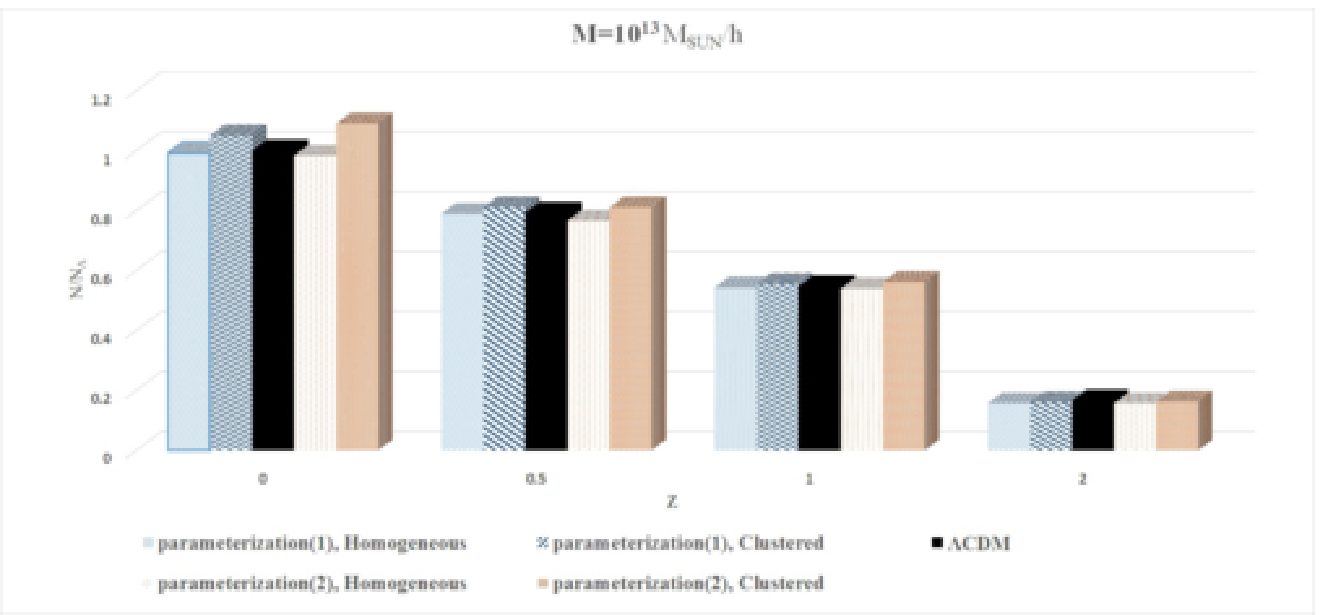}
	\includegraphics[width=0.87\textwidth]{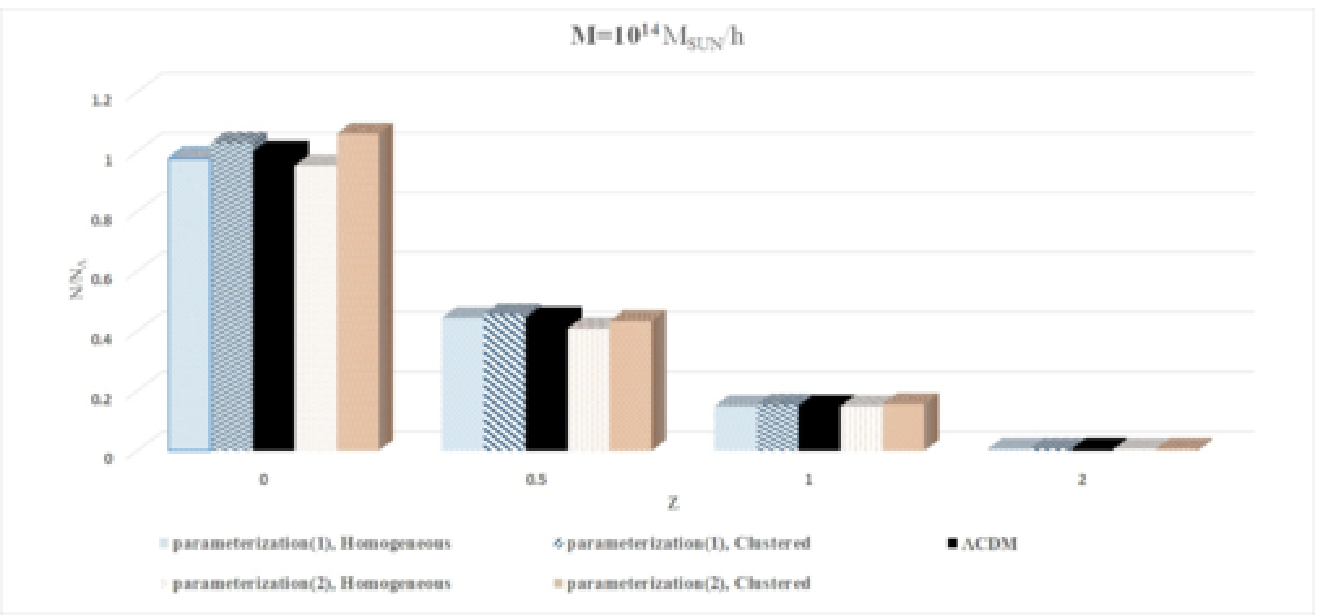}
    \includegraphics[width=0.87\textwidth]{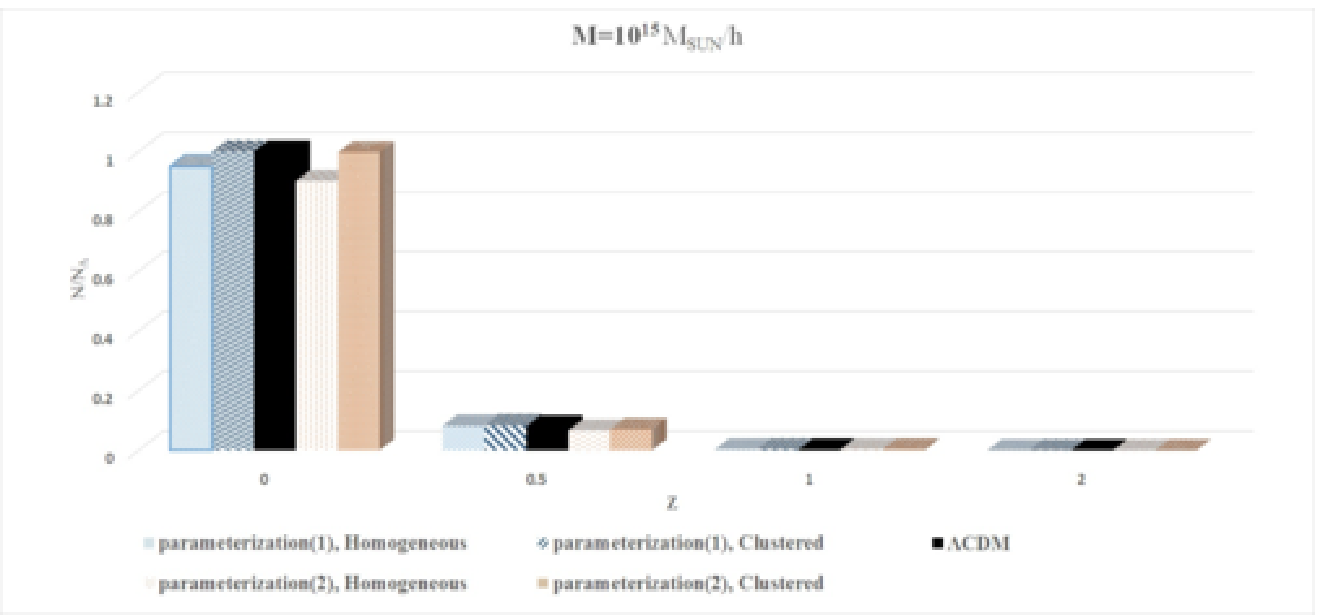}

	\caption{Value of number density of cluster-size halos normalized to that of the $\Lambda$CDM model at different redshifts calculated for different mass scales:$M>10^{13}M_{sun}h^{-1}$ (top panel) , $M>10^{14}M_{sun}h^{-1}$ (middle panel) and $M>10^{15}M_{sun}h^{-1}$ (bottom panel) for DE parameterizations in both clustered and homogeneous DE scenarios and $\Lambda$CDM model. Line styles and colors are shown in the legends. }
	\label{fig:nz}
\end{figure*}

At $z=0.5$, differences between various models are so small. Thus, in order to have better comparison, we use the numerical results from Table (\ref{tab:nZ0}), which contains precise numerical results of our analysis for three different mass scales. At this table, one can see that homogeneous DE scenario at $z=0.5$,led to less abundance of halos than the clustered DE  scenario. Clustered DE parameterizations, result somewhat more abundance of halos compared to homogeneous DE cases, while the difference is negligible at higher redshifts. In Fig. (\ref{fig:nz}) using the results presented in Table (\ref{tab:nZ0}), we visualize the predicted values of number density of halos normalized to that of the $\Lambda$CDM model. These values calculated for three different mass scales:$M>10^{13}M_{sun}h^{-1}$, $M>10^{14}M_{sun}h^{-1}$ and $M>10^{15}M_{sun}h^{-1}$. One can see that at all of mass scales, number density of halos decreases with increasing the redshift $z$. Moreover, comparing different panels of Fig. (\ref{fig:nz}) shows that for all models under study, the reduction of the number density of halos by $z$, in the case of massive halos is more significant. For example in the case of halos with mass higher than  $10^{15}M_{sun}h^{-1}$ for all of models, the ratio of $N/N_{\Lambda}(z=0)$ is $\sim3\times10^{-3}$ at $z=1.0$. While in the case of halos with mass higher than  $10^{14}M_{sun}h^{-1}$, these results obtain at $z=2$. These results shows that the massive dark matter halos were formed after smaller mass ones. Moreover, we can conclude that clustering of DE in both of parameterizations considered in this work, at all redshifts and in all mass scales, increases the number density of virialized halos. This effect is more significant at relatively low redshifts, where the universe is going to become DE dominated. This is expected, because these DE parameterizations were phantom at a wide range of redshifts. Same as the results of \citep{Abramo:2007iu} and our mentioned results in Sec.\ref{growth}, we know that clustering of DE in phantom models enhance the growth of dark matter perturbations and thus enhance the number density of virialized halos.

\begin{table*}
 \centering
 \caption{Ratio of the number of cluster-size halos above different value of mass $M$, for different DE parameterizations to the  $\Lambda$CDM model at $z=0$}
\begin{tabular}{c  c c c c c c }
\hline \hline
$z$ & $M[M_{sun}/h]$   &  $\Lambda$CDM  & \multicolumn{2}{|c}{   parameterization(1)   } &  \multicolumn{2}{|c}{   parameterization(2)  }\\
& &  &  Homogeneous  & Clustered  &  Homogeneous  & Clustered  \\
 \hline\\
&$10^{13}$ & 1.00 & 0.992 & 1.050 & 0.980 & 1.090 \\
 \\
$    z=0.0    $&$10^{14}$ & 1.00 & 0.977 & 1.030  & 0.954 & 1.060 \\
 \\
&$10^{15}$ & 1.00 & 0.951 & 1.00  & 0.902 & 0.997 \\
\hline\\
&$10^{13}$ & 0.793 &0.789 & 0.811 & 0.766& 0.811 \\
 \\
$z=0.5$&$10^{14}$ & 0.445 & 0.444 & 0.456  & 0.409& 0.434 \\
 \\
&$10^{15}$ & 0.080 & 0.082 & 0.084  & 0.066 & 0.070 \\
\hline\\
&$10^{13}$ & 0.547 & 0.543 & 0.555 & 0.540 & 0.562 \\
\\ 
$z=1.0$&$10^{14}$ & 0.150  & 0.149 & 0.153 &0.149 & 0.156\\
 \\
&$10^{15}$ & 3.0 E -3 & 3.1 E -3 & 3.2 E -3 & 3.2 E -3 & 3.5 E -3 \\
 \hline \\
&$10^{13}$ & 0.166 &0.157 & 0.160 & 0.157&0.162 \\
 \\
$z=2.0$&$10^{14}$ & 6.3 E -3 & 5.5 E -3 & 5.7 E -3  & 5.6 E -3 & 5.9 E -3 \\
 \\
&$10^{15}$ & 2.3 E -7  & 1.7 E -7 & 1.8 E -7  & 2.0 E -7& 2.2 E -7 \\

 \hline \hline
\end{tabular}\label{tab:nZ0}
\end{table*}

\section{CONCLUSION}\label{conclude}

In this work we studied the spherical collapse scenario in various DE parameterizations in which the EoS of DE is given by PADE approximation. We predicted number density of virialized halos for two DE parameterizations using Sheth-Tormen mass function. We first studied the evolution of Hubble expansion in selected parameterizations. We saw that the EoS parameter of parameterizations vary in phantom regime at high redshifts and cross the phantom line and enter in the quintessence regime at relatively low redshifts.
Then we investigated the effect of DE on the collapsing of dark matter halos in the  spherical collapse model. In particular, the effect of DE on the linear growth factor of perturbations, the linear and virial overdensities and  the abundance of dark matter halos was studied. 

Although, DE accelerates the expansion rate of the background of the universe, but it has two other different effects on the structure formation procedure. In the case of homogeneous DE, DE suppresses the growth of dark matter fluctuations. On the other hand, in the case of clustered DE, DE perturbations can enhance the growth of matter perturbations. Measuring the growth factor ,$D_+$, for all of DE parameterizations in both homogeneous and clustered DE scenarios results higher value compare to an EdS universe.

The important parameters of SCM,  $\delta_{\rm c}$ and $\Delta_{\rm vir}$ have been calculated for different parameterizations. We saw that in parameterization(1), the values of $\delta_{\rm c}$ and $\Delta_{\rm vir}$ are larger than values obtained for $\Lambda$CDM model. In the case of parameterization(2), these values are smaller than $\Lambda$CDM model. Also we saw that in the clustered cases of the both of parameterizations, in comparison with homogeneous cases, low dense virialized halos can be formed.

We also obtained the predicted number count of dark matter halos using the relevant Sheth-Tormen mass function for both of clustered and homogeneous DE scenarios respectively. In the case of clustered DE, by adding the contribution of DE mass on the total mass of clusters we applied the corrected form of mass function. 

We computed the number count of virialized halos at four different redshifts $z=0, 0.5, 1.0$ \& $2.0$. We saw that at all mentioned redshifts Sheth-Tormen mass function predicts more abundance of halos in clustered DE scenario compared to the homogeneous cases. We observe various results for the number count of halos in different DE parameterizations compare to the $\Lambda$CDM. Depending on the redshift $z$ and this fact that DE can be clustered or not, our results can be smaller or larger than that obtained in concordance $\Lambda$CDM cosmology.
Along the redshift, the number density of halos is decreasing. As expected, this decrement is more pronounced for massive halos compare to low mass halos. This result is compatible with this fact that the low mass  halos were formed before larger ones. We saw that the predicted number of dark matter halos in clustered DE cases is higher than that obtained in homogeneous DE cases. This means that in the DE parameterization under study, the clustering of DE component reinforces the formation of large scale structures. At high redshifts, $z=2$ where the abundance of halos falls down, the differences between different parameterizations and clustered and homogeneous approaches become negligible and we can not distinguish between them. Hopefully future observations of number count of cosmic structures can help us to distinguish among different scenarios of DE.
 
\section{Acknowledgements}
This work has been supported financially by Research Institute for Astronomy \& Astrophysics of Maragha (RIAAM) under research project No. 1/6025-33.

\bibliographystyle{mnras}
\bibliography{ref}
\label{lastpage}

\end{document}